\documentclass[hyper,12pt,letterpaper]{JHEP3}
\usepackage{epsfig,graphicx,bm,amsmath}
\newcommand{\be}{\begin{equation}}
\newcommand{\ee}{\end{equation}}
\newcommand{\bea}{\begin{eqnarray}}
\newcommand{\eea}{\end{eqnarray}}
\newcommand{\beq}{\begin{equation}}
\newcommand{\eeq}{\end{equation}}
\newcommand{\beqa}{\begin{eqnarray}}
\newcommand{\eeqa}{\end{eqnarray}}
\newcommand{\beqar}{\begin{eqnarray*}}
\newcommand{\eeqar}{\end{eqnarray*}}

\newcommand{\beas}{\begin{eqnarray*}}
\newcommand{\eeas}{\end{eqnarray*}}
\usepackage{epsf}

\title{
{\bf Moduli flow and non-supersymmetric AdS attractors}
}
\bigskip
\author{
 Dumitru Astefanesei$^{1}$\thanks{dumitru@th.phys.titech.ac.jp} ,
 Horatiu Nastase$^1$\thanks{nastase.h.aa@m.titech.ac.jp} ,
 Hossein Yavartanoo $^2$\thanks{yavar@phya.snu.ac.kr}
 ~and
 Sangheon Yun $^{3}$\thanks{sanhan1@phya.snu.ac.kr}\\
~\\
$^1${\it\small Global Edge Institute, Tokyo Institute of Technology, Tokyo 152-8550, JAPAN }\\
\\
$^2${\it\small Center for Theoretical Physics and BK-21 Frontier Physics Division,
 Seoul National University, Seoul 151-747 KOREA} \\
\\
$^3${\it\small School of Physics and Astronomy, Seoul National University,
 Seoul 151-747 KOREA  }\\}

\date{}

\maketitle
\newpage
\abstract{We investigate the attractor mechanism in gauged supergravity
in the presence of higher derivatives terms. In particular, we discuss
the attractor behaviour of static black hole horizons in anti-de Sitter
spacetime by using the effective potential approach as well as Sen's
entropy function formalism. We use the holographic techniques to interpret
the moduli flow as an RG flow towards the IR attractor horizon. We find that
the holographic c-function obeys the expected properties and point out some
subtleties in understanding attractors in AdS.}

\preprint{TIT/HEP-576}


\begin{document}
\vfill \setcounter{page}{0} \setcounter{footnote}{0}

\section{Introduction}
Black holes are testing grounds for string theory as a theory of quantum
gravity. The Bekenstein-Hawking entropy is inherently quantum
gravitational, involving both the Newtons's constant $G_N$ and Planck's
constant $\hbar$. Therefore, any consistent theory of quantum gravity
should address the origin of black hole entropy.

The `holographic' principle, \cite{'t Hooft:1993gx,Susskind:1994vu}, was formulated as an attempt at
understanding the physics of quantum black holes and at reconciling
gravitational collapse and unitarity of quantum mechanics at the Planck
scale. Thus, it is very tempting to consider the holographic principle as
a simple organizing principle for quantum gravity. String theory provides
a concrete realization of the holographic principle for spacetimes with negative
cosmological constant, namely the anti-de Sitter (AdS)/CFT correspondence \cite{Maldacena:1997re}.
That is a non-perturbative background independent
definition of quantum gravity in asymptotically AdS spaces. On the other hand, string
theory provides a microscopic description for the entropy of certain types of black holes
through the counting of D-branes bound states \cite{Strominger:1996sh, Maldacena:1996ky}.

AdS black holes in gauged supergravity
theories have found widespread application in the study of the AdS/CFT correspondence 
(see, e.g., \cite{Duff:1999rk} and references therein).
BPS objects are important within AdS/CFT duality, regardless of their precise nature, since
their properties remain the same in both the strong and weak coupling regimes of the
duality. However, it is also useful to investigate properties of non-BPS objects in
this context --- this is the topic of the present investigation.

The attractor mechanism plays a key role in understanding the entropy of
asymptotically flat non-supersymmetric extremal black holes in string theory 
\cite{Dabholkar:2006tb, Astefanesei:2006sy, Sen:2007qy}
and so it is of great interest to study the attractor behaviour of extremal black hole 
horizons in AdS.

The attractor mechanism was discovered in the context of $N=2$ supergravity
\cite{Ferrara:1995ih}, then
extended to other supergravity theories \cite{Ferrara:1996um}. 
It is now well 
understood that supersymmetry does not really play a fundamental role in
the attractor phenomenon. The attractor mechanism works as a consequence of the
symmetry of the near horizon extremal geometry that is given by $AdS_2\times S^p$
\cite{Sen:2005wa} for static spherically symmetric black holes
--- in fact, the `long throat' of $AdS_2$ (see \cite{klr})is at the basis of the attractor mechanism
\cite{Sen:2005wa, Astefanesei:2006dd, Kallosh:2006bt}.\footnote{A relation between the 
entanglement entropy of dual conformal quantum mechanics in $AdS_2/CFT_1$ and the 
entropy of an extremal black hole was provided in \cite{Azeyanagi:2007bj}.}

One can understand why the near horizon geometry is more important than 
supersymmetry by analogy with the flux compactifications: $AdS_2\times S^2$
can be interpreted as a flux compactification on $S^2$. This way, the flux generates 
an effective potential for the moduli such that, at the horizon, the potential has a stable 
minimum and the moduli are stabilized. Unlike the non-extremal case where the 
near horizon geometry (and the entropy) depends on the values of the moduli at 
infinity, in the extremal case the near horizon geometry is universal and is determined 
by only the charge parameters. Consequently, the entropy is also independent of the 
asymptotic values of the moduli. 

In this paper we study the attractor mechanism in AdS spacetime in the
presence of higher derivatives terms. We focus on static $5$-dimensional 
charged black hole solutions in gravity theories with $U(1)$  gauge fields 
and neutral massless scalars. The extremal black holes in AdS have also an  
$AdS_2 \times S^3$ geometry in the near horizon limit, hence the analogy 
would indicate that the attractor mechanism should also work  for this kind of 
black holes. 

Following \cite{Goldstein:2005hq}(for related work, see \cite{Alishahiha:2006ke}), we use 
perturbative methods and numerical analysis to show that the horizons of 
extremal black holes in AdS (with Gauss-Bonet term) are attractors --- this 
analysis supports the existence of the attractor mechanism for black holes in AdS 
space with higher derivatives. 

We will provide a physical interpretation for the attractor mechanism within 
the AdS/CFT duality. This requires the embedding in string theory that is explicitly 
constructed. Once we embed the solutions in 10 dimensional IIB supergravity 
(and so in string theory), we can use the AdS/CFT correspondence to interpret the moduli 
flow as a holographic renormalization group (RG) flow. 


To complete our analysis of the attractor mechanism within AdS/CFT duality, we will construct 
a c-function that obeys the expected results, namely it decreases monotonically as the radial 
coordinate is decreasing.  Therefore, within the AdS/CFT correpondence, there is a concrete connection between 
the attractor mechanism (gravity side) and the `dual' universality property of the QFT. The idea 
(reffered to as `universality' of QFT) that the IR  end-point of a QFT RG flow does not depend 
upon UV details becomes in the holography context the statement that the bulk solution for small 
values of $r$ does not depend upon the details of the matter at large values of $r$. 
Indeed, within the attractor mechanism, the black hole horizon (IR region) does not have any 
memory of the initial conditions (the UV values of the moduli) at the boundary. The black hole 
entropy depends just on the charges and not on the asymptotic values of the moduli. However, 
we can interpret it as a `no-hair' theorem for the extremal black holes in AdS that is equivalent 
with the `universality'  of the field theory on the brane \cite{Warner:2002ip}.

The  paper is organized as follows: In section 2 we discuss the attractor mechanism in two
derivatives gauged supergravity. Our discussion is general in that it is based on analysis of
the equations of motion, not just the near horizon geometry and its symmetries. We show the
equivalence of the effective potential approach \cite{Goldstein:2005hq} and the entropy
function formalism \cite{Sen:2005wa} in the near horizon limit of the extremal black
holes in AdS. In section 3 we examine the attractor mechanism in AdS gravity with higher
derivatives. We generalize the effective potential in the Gauss-Bonnet gravity and find that
even in this case the extremal black hole horizon is a stable minimum of the effective potential.
Consequently, the moduli are stabilized and the entropy does not depend on couplings.
In section 4 we present a holographic interpretation for the attractor mechanism
by identifying the moduli flow with the RG flow and also find the c-function. Finally, we end with a 
discussion of our results in section 5.

\section{Attractors in two derivatives gauged supergravity}
In this section we generalize the results of \cite{Goldstein:2005hq} by
including a potential for the scalar fields in the action. We discuss
the attractor mechanism using both the effective potential method
\cite{Goldstein:2005hq} and the entropy function framework
\cite{Sen:2005wa}. The first method is based on investigating
the equations of motion of the moduli and finding the conditions
satisfied by the effective potential such that the attractor
phenomenon occurs. The entropy function approach is based on the
near horizon geometry and its enhanced symmetries.

\subsection{Generalities}
While details of the various supergravity theories depend crucially on the dimension,
general features of the bosonic sector can be treated in a dimension independent
manner. However, from now on, we will focus on a 5-dimensional theory of gravity coupled
to a set of massless scalars and vector fields, whose general bosonic action has the
form
\bea
  \nonumber
  I[G_{\mu\nu},\phi^i,A_{\mu}^I]
  \!&=&\!\frac{1}{\kappa ^2}\int_{M} d^{5}x
  \sqrt{-G}[ R-g_{ij}(\phi)\partial_\mu\phi^i\partial^\mu\phi^j \\
  &&-f_{AB}(\phi)F^{A}_{\mu\nu}F^{B\, \mu\nu}+V(\phi)]
  \label{actiongen}
\eea
where $F^A_{\mu\nu}$ with $A=(0, \cdots N)$ are the gauge fields, $\phi\equiv (\phi^i)$
with $i=(1, \cdots, n$) are the scalar fields, $V(\phi^i)$ is the scalar fields
potential, and $\kappa^2=16\pi G_N$. The moduli determine the gauge coupling constants
and $g_{ij}(\phi)$ is the metric in the moduli space. We use Gaussian units so
that factors of $4\pi$ in the gauge fields can be avoided and the
Newton's constant $G_N$ is set to $1/16\pi$. The above action is of the type of the $gauged$
supergravity theories.\footnote{In 5-dimensional supergravity theories, one should also
consider a gauge Chern-Simons term. However, since we are considering only static
electrically charged black hole solutions, the Chern-Simons term does not play any role.}

The equations of motion for the metric, moduli, and the gauge fields are given by
\bea
  \label{einstein}
  R_{\mu\nu}-g_{ij}\partial_{\mu}\phi^i\partial_{\nu}\phi^j
  = f_{AB}\left(2F^A_{\phantom{A}\mu\lambda}F^{B\phantom{\nu}\lambda}_
    {\phantom{B}\nu}-
    {\textstyle \frac{1}{3}}G_{\mu\nu}F^A_{\phantom{a}\alpha\lambda}
    F^{B \alpha\lambda} \right) -\frac{1}{3}G_{\mu\nu} V(\phi)\, 
\eea
\bea
  \frac{1}{\sqrt{-G}}\partial_{\mu}(\sqrt{-G}g_{ij}\partial^{\mu}\phi^j)
  = \frac{1}{2}\left(\frac{\partial f_{AB}}{\partial \phi^i} F^A _{\phantom{A}\mu\nu}
  F^{B\, \mu\nu} + \frac{\partial g_{mn}}{\partial \phi^i}\partial_{\mu}\phi^m\partial^{\mu}\phi^n - \frac{\partial V(\phi)}{\partial \phi^i}\right) \, 
  \label{dilaton}
\eea
\bea
  \partial_{\mu}\left[\sqrt{-G}\left(f_{AB} F^{B\, \mu\nu}\right)  \right] =  0\, 
  \label{gaugefield}
\eea
where we have varied the moduli and the gauge fields independently. The Bianchi identities
for the gauge fields are $F_{\phantom{A}\,[\mu\nu;\lambda]}^{A}=0$.\footnote{From now on
we keep the metric on the moduli manifold constant --- the conditions for the existence
of attractor mechanism will not change if we allow a moduli dependence for the metric.}

We focus on $5$-dimensional spherically symmetric spacetime metrics and we consider the
following ansatz:
\bea
  ds^{2} & = & -a(r)^{2}dt^{2}+a(r)^{-2}dr^{2}+b(r)^{2}d\Omega_3^{2}
  \label{metric2}
\eea
We consider a definite form of the 3-sphere
\bea
d\Omega^2_3 &=& d\theta^2 + \sin^2\theta\, d\phi^2
+\cos^2\theta\, d\psi^2\, 
\eea
with coordinate ranges $\theta \in [0, \frac{\pi}{2}]$ and
$\phi,\psi \in [0, 2\pi]$.

The Bianchi identity and equation of motion for the gauge fields can be solved by a field
strength of the form \cite{Goldstein:2005hq}
\bea
  \label{fstrenghtgen}
  F^A= \frac{1}{b^3} f^{AB}Q_{B} \; dt\wedge dr
\eea
where $Q_{A}$ are constants which determine electric charges carried
by the gauge field $F^A$ and $f^{AB}$ is the inverse of $f_{AB}$.

With this ansatz, the gravitational equations
of motion become
\bea
R_{rr} &=& - \frac{1}{a^2b} \left( b(a'^2 + aa'')+ 3 a (a'b' + ab'') \right) \\
\nonumber
    &=&  \phi'^2 - \frac{4}{3a^2b^6}\, f^{AB}Q_AQ_B -\frac{1}{3a^2}\, V(\phi)\,  \\
R_{tt} &=& a^2 (a'^2 + \frac{3aa'b'}{b} + a a'')\\
\nonumber
    &=& \frac{4 a^2}{3 b^6} f^{AB}Q_AQ_B + \frac{1}{3} a^2 V(\phi)
  \, \\
R_{\theta\theta}&=& 2 -2 aba'b' -a^2(2b'^2+bb'')\\
\nonumber
&=&\frac{2}{3b^4} f^{AB}Q_AQ_B -\frac{1}{3}b^2 V(\phi)
\eea
Here we use the notation $\phi'^2=g_{ij}\partial_r\phi^i\partial^r\phi^j$.
Note that $R_{\phi\phi}=\sin^2\theta R_{\theta\theta},~
R_{\psi\psi}=\cos^2\theta R_{\theta\theta},$ and that off-diagonal
components of the Ricci and stress tensors vanish. It is also important to notice 
that the field equations are not all independent.

It is easier to use combinations of the equations above
\beqa
R_{tt} + 2 \frac{G_{tt}}{G_{\theta\theta}} R_{\theta\theta}\, , \hspace{0.5cm}
R_{rr} - \frac{G_{rr}}{G_{tt}} R_{tt}\, , \hspace{0.5cm}
-\frac{G_{\theta\theta}}{G_{tt}} R_{tt} + \frac{G_{\theta\theta}}{G_{rr}} R_{rr}  - 3 R_{\theta\theta}
\eeqa
and from now on we will work with the following equivalent system
of differential equations:
\bea
\label{eqab}
0&=&4(-1+a^2b'^2) + (a'^2+aa'')b^2+ ab(7a'b' + 2ab'') - b^2 V(\phi_i)\\
\label{eqphi}
0&=&\phi'^2 + 3 \frac{b''}{b}\\
\label{const}
0&=&-1+aba'b' + a^2b'^2-\frac{1}{6} a^2 b^2 \phi'^2 - \frac{1}{6} \; b^2 V(\phi_i) + \frac{V_{eff}}{3 b^4}
\eea
We should also consider the equations of motion for the scalars which can be written as
\beq
\label{phi}
\partial_r (a^2 b^3 \partial_r \phi_i) = \frac{1}{ b^3} (\partial_i V_{eff} - \frac{1}{2} b^6 \partial_i V)
\eeq
where $V_{eff}= f^{AB} Q_A Q_B$ and $f^{AB}$ is the inverse of $f_{AB}$. When the scalar
potential $V(\phi)$ is constant, $V_{eff}(\phi^i)$ plays the role of an `effective
potential' that is generated by non-trivial form fields. The effective potential, first
discussed in \cite{Ferrara:1997tw}, plays an important role in describing the
attractor mechanism \cite{Goldstein:2005hq, Tripathy:2005qp}.

A vanishing Hamiltonian is a characteristic feature of any theory which is invariant
under arbitrary coordinate transformations --- for our system, the equation (\ref{const})
does not contain any second derivatives and is the Hamiltonian constraint.

As a final comment, we observe that the equations of motion can also be obtained from
the following one-dimensional action
\beq
S= \frac{1}{\kappa^2} \int dr \left(6b + 6a b^2 a' b' + 6 a^2 b b'^2 + b^3 V(\phi) - 
a^2 b^3 (\phi_i')^2 -\frac{2}{b^3} V_{eff}(\phi_i) \right)
\eeq

\subsection{Entropy function}
We apply the entropy function formalism to static black holes in AdS
space.\footnote{The entropy function for AdS black holes was considered
by Morales and Samtleben in \cite{Morales:2006gm}. However, our discussion is 
more general and
the interpretation of some results in this section are substantially
new.} It was shown by Sen that the attractor mechanism is related to the
extremality (attempts to apply the entropy function to non-extremal black holes can 
be found in \cite{Cai:2007ik}) rather than to the supersymmetry property of a given solution.
Indeed, the $AdS_2$ factor of the near-horizon geometry is at the basis
of the attractor mechanism. As has been discussed in \cite{Ferrara:1997yr,
Kallosh:2006bt}, the moduli do not preserve any memory of the initial
conditions at infinity
due to the presence of the infinite throat of $AdS_2$. This is in analogy
with the properties of the behaviour of dynamical flows in dissipative
systems, where, on approaching the attractors, the orbits
practically lose all the memory of their initial conditions.\footnote{This analogy 
should be taken with caution --- a detailed discussion on this subject can be 
found in \cite{Astefanesei:2006sy}.}

Therefore, an important hint for the existence of the attractor meachanism
is the existence of an $AdS_2$ as part of the near horizon geometry of an
extremal black hole. The extremal charged black hole solution of the equations of
motion with constant scalar fields is the extremal Reissner-Nordstrom-anti-de Sitter
(RNadS) black hole given by \cite{Cardoso:2004uz}
\bea
\label{rnads}
a^2(r)=1+\frac{r^2}{l^2}-\frac{m}{r^2}+\frac{q^2}{r^4}=
\frac{1}{l^2r^4}(r-r_H)^2(r+r_H)^2(r^2+2r_H^2+l^2)
\eea
Here $r=r_H$ is the degenerate horizon and can be calculated using
the following expressions of mass and charge parameter:
\beq
m=2r_H^2\left(1+\frac{3}{2}\frac{r_H^2}{l^2}\right), \hspace{0.6cm}
q^2=r_H^4\left(1+2\frac{r_H^2}{l^2}\right)
\eeq
The mass parameter $m$ and the charge parameter $q$ are related to the
asymptotic ADM charges $M$ and $Q$ by:
\bea
\label{ADM}
M=\frac{3\pi}{8G_N}m\, , \hspace{1cm} Q=\sqrt{3}q
\eea
and the electric field is given by
\beq
F=\frac{1}{2}F_{\mu\nu}dx^{\mu}\wedge dx^{\nu}=\frac{Q}{r^3}dr\wedge dt\, 
\eeq
In the near horizon limit, $\rho=r-r_H\rightarrow 0$, we obtain
\beq
a(\rho)=\frac{4}{l^2r_H^2}(3r_H^2+l^2)\, \rho^2=\frac{1}{v_1}\, \rho^2
\eeq
where $v_1$ is a constant that can be interpreted as the radius of
$AdS_2$ --- the $AdS_2\times S^3$ geometry appears explicitly by
making the change of coordinates $t=v_1\tau$.

It is important to notice that the extremal solution is non-supersymmetric. The
supersymmetric bound is $M=2Q$ and in this limit one finds a naked curvature
singularity at $r=0$. However, by adding $\alpha'$-corrections this singularity
may be dressed by a horizon with finite area.

Let us now briefly review the entropy function formalism. In \cite{Sen:2005wa} 
(see also \cite{Cai:2007an}), it
was observed that the entropy of a spherically symmetric extremal
black hole is the Legendre transform of the Lagrangian density.  The derivation of this
result does not require the theory and/or the solution to be supersymmetric. The only
requirements are gauge and general coordinate invariance of the action.

The entropy function is defined as
\bea
  F(\overrightarrow{u},\overrightarrow{v},\overrightarrow{e},
  \overrightarrow{p})=
  2\pi(e_iq_i-f(\overrightarrow{u},\overrightarrow{v},
  \overrightarrow{e},\overrightarrow{p})=
  2\pi(e_iq_i-\int d\theta d\phi d\psi \sqrt{-G}{\cal L})
\eea
where $q_i=\partial f/\partial e_i$ are the electric charges, $u_s$ are the values of the
moduli at the horizon, ${p_i}$ and ${e_i}$ are the near horizon radial magnetic and
electric fields and $v_1$, $v_2$ are the sizes of $AdS_2$ and $S^2$ respectively. Thus,
$F/2\pi$ is the Legendre transform of the function $f$
with respect to the variables $e_i$.\footnote{The reason why it is not a Legendre
transform with respect to magnetic charges is due to topological character of the
magnetic charge. The Bianchi identities do not change when the action is supplemented
with $\alpha'$-corrections, but the equations of motion receive corrections.}

For an extremal black hole of electric charge $\overrightarrow{Q}$ and magnetic charge
$\overrightarrow{P}$, Sen has shown that the equations determining
$\overrightarrow{u},\overrightarrow{v}$, and $\overrightarrow{e}$ are given by:
\bea
  \frac{\partial F}{\partial u_s}=0\,, \qquad
  \frac{\partial F}{\partial v_i}=0\,,
  \qquad \frac{\partial F}{\partial e_i}=0\,
  \label{attractor}
\eea
Then, the black hole entropy is given by
$S=F(\overrightarrow{u},\overrightarrow{v},\overrightarrow{e}, \overrightarrow{p})$ at the
extremum (\ref{attractor}). The entropy function,
$F(\overrightarrow{u},\overrightarrow{v},\overrightarrow{e}, \overrightarrow{p})$,
determines the sizes $v_1$, $v_2$ of $AdS_2$ and $S^3$ and also the near horizon values of
moduli ${u_s}$ and gauge field strengths ${e_i}$. If $F$ has no flat directions, then the
extremization of $F$ determines $\overrightarrow{u}$, $\overrightarrow{v}$,
$\overrightarrow{e}$ in terms of $\overrightarrow{Q}$ and $\overrightarrow{P}$. Therefore,
$S=F$ is independent of the asymptotic values of the scalar fields. These results lead to
a generalised attractor phenomenon for both supersymmetric and non-supersymmetic extremal
black hole solutions.

Now we are ready to apply this method to our action (\ref{actiongen}).
The general metric of $AdS_2\times S^3$ can be written as
\bea
  ds^2=v_1(-\rho^2d\tau^2+\frac{1}{\rho^2}d\rho^2)+
  v_2d\Omega_3^2\, 
  \label{adshor}
\eea
The field strength ansatz (\ref{fstrenghtgen}) in our case is given by
\bea
  F^A=e^Ad\tau\wedge d\rho.
\eea
The entropy function $F(u^i,v_1, v_2, e^A, Q_A)$ and $f(u^i,v_1, v_2, e^A)$ are given by
\bea
\label{Fmisto}
  && F(u^i,v_1,v_2,e^A,Q_A)=2\pi [Q_Ae^A-f(u^i,v_1, v_2, e^A)]\, ,\\
  \nonumber
  && f(u^i,v_1, v_2, e^A)=2\pi^2\left[-2v_2^{3/2}+6v_1\sqrt{v_2}+
    2\frac{v_2^{3/2}}{v_1}f_{AB}e^Ae^B+v_1v_2^{3/2}V(\phi)\, \right]\, 
\eea
Then the attractor equations are obtained as :
\bea
  \label{at1}
  \frac{\partial F}{\partial v_1} & = & 0\,\,\,\Rightarrow
  \,\,\,6v_1^2-2v_2f_{AB}e^Ae^B+v_1^2v_2V(\phi)=0   \, \\
  \label{at2}
  \frac{\partial F}{\partial v_2} & = & 0\,\,\,\Rightarrow \,\,\,-v_1v_2+
v_1^2+v_2f_{AB}e^Ae^B+\frac{v_1^2v_2}{2}V(\phi)=0\, \\
  \label{at3}
  \frac{\partial F}{\partial u^i} & = & 0\,\,\,\Rightarrow \,\,\,
   \,2\frac{\partial f_{AB}}{\partial u^i}e^Ae^B=-v^2_1\frac{\partial V}{\partial u^i} \\
  \label{at4}
  \frac{\partial F}{\partial e^A} & = & 0\,\,\,\Rightarrow \,\,\,
  Q_A=8\pi^2\, \frac{v_2^{3/2}}{v_1}f_{AB}e^B\, 
\eea
By combining the first two equations we obtain $4/v_2-1/v_1+V(\phi)=0$ and so the radii
of $AdS_2$ and $S^3$ are related by the potential of the scalars.\footnote{We can check
this relation for the extremal RNadS black hole \cite{Cardoso:2004uz} by using the
following relations: $V(\phi)=-4\Lambda$, $\Lambda=-3/l^2$, $v_2=r_H^2$, and $v_1$ is
given in (\ref{rnads}).}  By replacing (\ref{at1}) and (\ref{at4}) in (\ref{Fmisto}) we
obtain the value of the entropy function at the extremum, $F=8\pi^3v_2^{3/2}$, that
is the entropy of the black hole (our convention was $G_N=1/16\pi$).

The third equation is very important: in AdS spacetime, $V(\phi)=constant$, this
equation is equivalent with finding the critical points of the effective potential
at horizon. One can easily eliminate the field strengths in the favour of charges
by using the last equation to obtain
$(\partial f^{AB}/ \partial u^i) \, Q_AQ_B=0$ --- we will show in the next subsection
that this is one of the conditions for the existence of attractor mechanism. If this
equations has solutions, then the moduli values at the horizon are fixed in term of
the charges. It is also important to notice that the existence of a near-horizon
geometry when the moduli are not constants does not imply the existence of the
whole solution in the bulk (from the horizon to the boundary) --- this is the disadvantage
of the entropy function formalism. However, in the next subsection we will investigate
the equations of motion in the bulk and describe the horizon as an IR critical
point of the effective potential.

\subsection{Effective potential and non-supersymmetric attractor}
\label{veff}
In this section we consider a {\it constant} potential for scalars,
$V(\phi)=12/l^2$. For
the attractor phenomenon to occur, it is sufficient if the following two
conditions are satisfied \cite{Goldstein:2005hq}. First, for fixed charges,
as a function of the moduli, $V_{eff}$ must have a critical point. Denoting
the critical values for the scalars as $\phi^i=\phi^i_0$ we have,
\bea
  \label{critical}
  \partial_iV_{eff}(\phi^i_{0})=0
\eea
Second, there should be no unstable directions about this minimum, so the matrix of second
derivatives of the potential at the critical point,
\bea
  \label{massmatrix}
  M_{ij}=\frac{1}{2} \partial_i\partial_jV_{eff}(\phi^k_{0})
\eea
should have no negative eigenvalues.  Schematically we can write,
\bea
  \label{positive}
  M_{ij}>0
\eea
We will refer to $M_{ij}$ as the mass matrix and its eigenvalues as masses (more
correctly $mass^2$ terms) for the fields, $\phi^i$.

It is important to note that in deriving the conditions for the attractor phenomenon, one
does not have to use supersymmetry at all. The extremality condition puts a strong
constraint on the charges so that the asymptotic values of the moduli do not appear in the
entropy formula.

\subsubsection{Zeroth order analysis }
Let us start by setting the asymptotic values of the scalars equal to
their critical values (independent of $r$), $\phi^i=\phi^i_0$. The
equations of motion (\ref{eqphi}, \ref{eqab}) can be easily solved.
First we solve (\ref{eqphi}) and get $b(r)=r$, and then replace this
expression in (\ref{eqab}) --- we obtain:
\beqa
\frac{1}{2}r^2(a^2)''+\frac{7}{2}r(a^2)'+4a^2=4+\frac{12}{l^2}r^2.
\eeqa
The most general solution of this equation is given by
$a^2(r)=1+C_1/r^2+C_2/r^4+r^2/l^2$, where $C_1$ and $C_2$
are integration constants. We are interested in the extremal solutions
and so the integration constants can be calculated from the `double horizon'
\footnote{The inner and outer horizons coincide and the equation has a
double root.} condition:
\beq
C_2= - (\frac{3r_H^4}{l^2} + 2r_H^6), \;\;\;\;\; C_1= r_H^4 + \frac{2r_H^6}{l^2},
\eeq
where $r_H$ is the horizon radius.
Therefore, we can write the solution as
\beq
\label{a0}
a_0(r) = (1-\frac{r_H^2}{r^2})\sqrt{1+ \frac{r^2+ 2r_H^2}{l^2}} , \;\;\;\ b_0(r)=r,
\eeq
that describes the extremal RNadS found in the previous subsection.

The Hamiltonian constraint evaluated at the boundary provides a constraint
on charges. However, we are interested in solving the Hamiltonian constraint
at the horizon and to obtain a relation between the entropy and the effective
potential. It is important to notice that the temperature is proportional
to $aa'$ and so {\it just} in the extremal limit this product is vanishing. With
this observation the Hamiltonian constraint simplifies drastically at the horizon.
Thus, the horizon radius, $r_H$, can be computed from the following equation:
\beq
\label{horizonR}
-3r_H^4-\frac{6}{l^2}r_H^6+V_{eff}(\phi^i_0)=0.
\eeq
We obtain ($f_{AB}=1$)
\beq
Q^2=3r_H^4\, \left(1+2\frac{r_H^2}{l^2}\right),
\eeq
that is the $ADM$ electric charge (\ref{ADM}) of the extremal RNadS black hole.

\subsubsection{First order analysis}
For the extremal RNadS black hole solution carrying the charges specified by
the parameter $Q_A$ and the moduli taking the critical values $\phi^i_{0}$ at
infinity, a double zero horizon continues to exist for small deviations from these
attractor values for the moduli at infinity. The moduli take the critical values at the
horizon and entropy remains independent of the values of the moduli at infinity
\cite{Goldstein:2005hq}. The horizon radius is given by the eq.~(\ref{horizonR})
and the entropy is
\bea
  \label{BH}
  S_{BH}=\frac{A}{4G_N}= \frac{\pi^2}{2G_N} r_H^3=8\pi^3 r_H^3
\eea

We start with first order perturbation theory
\beq
\label{phi1}
\delta \phi_i = \phi_i -\phi_{i0} = \epsilon \phi_{i1}
\eeq
where $\epsilon $ is a small parameter we use to organize the perturbation theory.  The first correction to the scalars $\phi_i$ satisfies the equation
\beq
\label{phi2}
\partial_r (a_0^2 b_0^3 \partial_r \phi_{i1}) = \frac{ \beta_i^2}{ b_0^3} \phi_{i1}
\eeq
where $\beta_i^2$ is the eigenvalue for the matrix $2 M_{ij}$. We are interested in a `smooth' solution that does not
blow up at horizon $r=r_H$. It is difficult to find a general solution --- however we will study our equations in the
near horizon limit (the solution in the asymptotic region is presented in Section 4) and keep in mind that there is a
smooth interpolation between the horizon and the boundary. In the near horizon limit, we obtain
\be
\label{phi3}
\phi_{i1} = c_{1i} (1-\frac{r_H}{r})^{\gamma_i}
\eeq
where $\gamma_i$ are positive roots of following equations
\beq
\label{gamma1}
\gamma_i (\gamma_i+1) = \frac{\beta_i^2}{4r_H^4} (1+ \frac{3r_H^2}{l^2})^{-1}
\eeq
Asymptotically (as $r\rightarrow \infty$) $\phi_{i1}$ takes a constant value, $c_{1i}$ --- however $\phi_{i1}$ is vanishing
at the horizon and the value of the scalar is fixed at $\phi_{i0}$ regardless of its value at infinity.
We observe from the equation (\ref{gamma1}) that if the eigenvalues of the mass matrix are positive, 
then the solution is regular at the horizon and so the existence of a regular horizon is related to 
the existence of the attractor mechanism. In the light of previous discussions, this is easy to understand 
if we recall that the near horizon geometry of an extremal black hole is $AdS_2\times S^3$.

\subsubsection{Second order analysis and back reaction}

The first perturbation in scalars sources a second order correction in the metric. We write
\bea
a&=& a_0 + a_2 \epsilon^2 \\
b&=& b_0 + b_2 \epsilon^2
\eea
and by solving the equations (\ref{eqab}) and (\ref{eqphi}) we obtain
\bea
a(r) &=& (1-\frac{r_H^2}{r^2}) \sqrt{1+\frac{r^2+ 2r_H^2}{l^2}} \left(1+a_{i2} (1-\frac{r_H}{r})^{2\gamma_i}\right) \\ \cr
 b(r) &=& r \left(1+b_{i2} (1-\frac{r_H}{r})^{2\gamma_i}\right)
\eea
where
\bea
a_{i2} &=& - b_{i2} \left(\frac{1}{(\gamma_i+1)(2\gamma_i+1)(1+\frac{3r_H^2}{l^2})} + \frac{\gamma_i(4\gamma_i+5)}{(\gamma_i+1)(2\gamma_i+1)}\right) \\
b_{i2} &=& -\frac{\gamma_i c_{1i}^2}{6(2\gamma_i-1)}
\eea
We see that in second order we need to choose again positive $\gamma_i$ in order  to get a regular 
horizon. That means the small fluctuations about the extremal point must all be 
positive and so the horizon is an attractor. Thus, in the near horizon limit we obtain 
again the near horizon geometry of the extremal RNadS black hole that is fixed only 
by the charges.

\subsection{Higher order result}
Going to higher orders in perturbation theory is in principle straightforward. We solve the system of equations (\ref{eqab})-(\ref{const}) order by order in the $\epsilon$-expansion. To first order, we find that one variable, say $c_{1i}$ , can not be fixed by the equations.
Thus we find $a_{i2}$ and $b_{i2}$ as functions
of $c_{1i}$. One can check that at any order $n > 2$, one can substitute the resulting values
of ($a_{mi} , b_{mi} , \phi_{mi}$ ), for all $m \leq n$ from the previous orders.                                                             Then (\ref{eqab})-(\ref{const}) of the  order $m$, consistently give,
\be
a_{mi}= a_{mi}(c_{1i}), \;\;\;\; b_{mi}= b_{mi}(c_{1i}), \;\;\;\; \phi_{mi}= \phi_{mi}(c_{1i}),
\ee
as polynomials of order n in terms of $c_{1i}$. It is worth noting that $c_{1i}$ remains a free parameter to all orders in the $\epsilon$-
expansion. Owing to the result above, we observe that $(a_{\infty i} , b_{\infty i} , 
\phi_{\infty i} )$ are varying and will take
different values, given different choices for  $c_{1i}$. The arbitrary value of $\phi_{i}$ at infinity is $\phi_i=\phi_{\infty i}$ , while its value at the horizon is fixed to be $\phi_{0i}$. Figure 1 shows 
the result of numerical simulations for $\phi$ vs. $r$ with different asymptotic values $\phi_{\infty}$.

\begin{figure}
 \begin{center}
  \includegraphics[scale=.6]{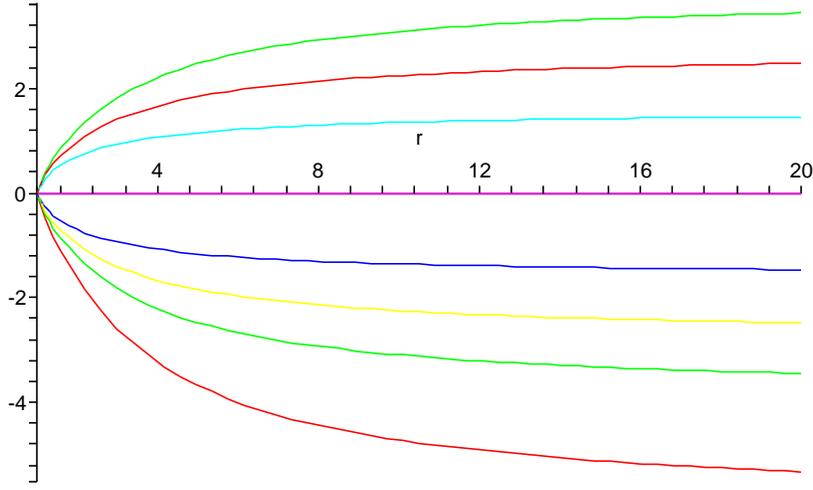}
 \end{center}
\caption{$\phi(r)$ vs. $r$, where the 
numerical coefficients are $r_H=1, l^2=6$ for the effective potential $V_{eff} = 2 e^{\sqrt{3}\phi} + 2 e^{-\sqrt{3} \phi}$. Different curves represent different asymptotic values for $\phi_{\infty}$. The attractor point is $\phi_0=0$ at the horizon,  $r_H=1$.}
\end{figure}

\begin{figure}
 \begin{center}
  \includegraphics[scale=.6]{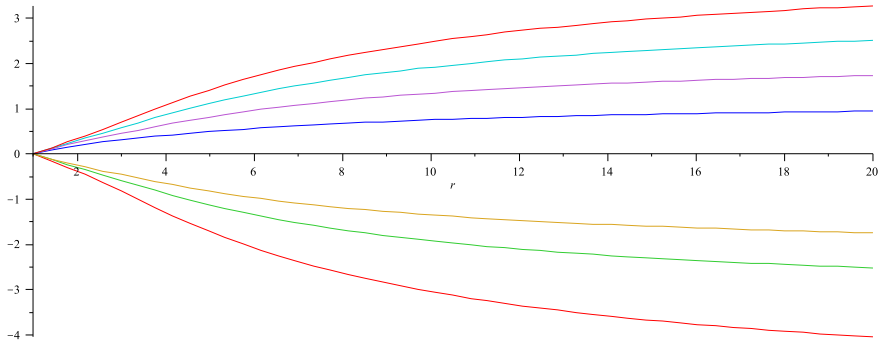}
 \end{center}
\caption{$\phi(r)$ vs. $r$, where the 
numerical coefficients are $r_H=1, l^2=6$ for the effective potential $V_{eff} = 2 e^{3\phi} + 2 e^{-3 \phi}$. Different curves represent different asymptotic values for $\phi_{\infty}$. The attractor point is $\phi_0=0$ at the horizon,  $r_H=1$.}
\end{figure}

\section{AdS attractors with higher derivatives}
The process of compactification of the string theory from higher to lower dimensions introduces
scalar fields (moduli/dilaton) which are coupled to curvature invariants. We prove the existence 
of the attractor mechanism even in the presence of  higher derivatives terms. For simplicity, we 
consider just $R^2$ corrections which appear in bosonic string theory but we expect to reach
similar conclusions for more interesting case of the  $R^4$ corrections.

\subsection{Equations of motion for Gauss-Bonnet gravity}
We add the most general $R^2$ correction with general scalar coupling to our previous action. The action is given by:
\beq
S=S_0 + S_{corr}
\eeq
where
\bea
S_{corr} = \frac{1}{\kappa^2}\int d^5x \sqrt{-g} \bigg[ G_1(\phi_i) R^2 + G_2(\phi_i) R_{\mu\nu} R^{\mu\nu} + G_3(\phi_i) R_{\mu\nu\alpha\beta} R^{\mu\nu\alpha\beta} \bigg] \nonumber
\eea
From now on we focus on $R^2$ corrections which form the Gauss-Bonnet Lagrangian\footnote{The Gauss-Bonnet term is in 
fact the most general ghost-free combination of $R^2$ terms in any dimension and is a total derivative up to 
order $h^2$ in $h_{\mu\nu}$  \cite{zwiebach}, so is a good primer of quantum gravity corrections.}
\beq
L_{GB}=R^2-4R_{\mu\nu}R^{\mu\nu}+R_{\mu\nu\alpha\beta}R^{\mu\nu\alpha\beta}.
\eeq
that correspond to $G_2 = - 4 G_1, G_3 =G_1$ in the above most general action. 
The equations of motion for the gauge fields do not change, while the scalar and three Einstein equations are modified by the Gauss-Bonnet term. Now we are again interested in static, spherically symmetric black holes. Thus, we
consider the same ansatz as in the previous section for the gauge fields and the following ansatz for the metric:
\be
  ds^{2} = -a(r)^{2}dt^{2}+c(r)^{-2}dr^{2}+b(r)^{2}d\Omega_3^{2}
\ee
Then all the equations can be obtained from the following one-dimensional action (see the appendix for details):
\bea
S_{1-dim} = &&\frac{\pi}{4G_N}\int dr \bigg[ 3b^2b'a'c + \frac{3ab}{c}(1-b'^2c^2) + 6abcb'^2 - \frac{1}{2}ab^3c \phi'^2 - \frac{a}{b^3c}V_{eff}(\phi) + \frac{6ab^3}{l^2 c} \cr && + 12G(\phi)(ab'^3c^2c' + a'b'c + ab'^2b''c^3) -12G'(\phi)(a'bb'^2c^3 - a'bc - ab'c) \bigg]
\eea
After a little algebra, choosing the gauge $a=c$, the Einstein equations can be written as
\bea
\label{eqgb1} &&\phi'^2 + 3 \frac{b''}{b^3} \left( b^2+ 4 G(\phi_i) (1-a^2b'^2) - 8G'(\phi_i)a^2bb' \right) + \frac{12G''(\phi_i)}{b^2}(1-a^2b'^2)= 0 \\ \cr
\label{eqgb2}&&4(-1+a^2b'^2) + (a'^2+aa'')b^2+ ab(7a'b' + 2ab'')  - \frac{12 b^2}{l^2} \cr
&& + \frac{4G(\phi_i)}{b} {\mathcal G}_1 - \frac{4G'(\phi_i)}{b} {\mathcal G}_2 + 4G''(\phi_i)(a^2-a^4b'^2-2a^3a'bb')=0 \\ \cr
\label{eqgb3}&&-1+aa'bb' + a^2b'^2 - \frac{1}{6}a^2b^2 \phi'^2 + \frac{V_{eff}}{3 b^4} - \frac{2 b^2}{l^2} + \frac{4G(\phi_i)}{b} (aa'b' -a^3a'b'^3) \cr
&& - \frac{4G'(\phi_i)}{b} (3a^3a'bb'^2 - aa'b - a^2b' + a^4b'^3) =0
\eea
where
\bea
{\mathcal G}_1 &=& (1-a^2b'^2)(3aa'b'+aa''b) + a'^2b(1-3a^2b'^2)-2a^3a'b b'b'' \\ 
{\mathcal G}_2 &=& 6a^2a'^2b^2b' - 3a^2b'(1-a^2b'^2)- 5aa'b(1-3a^2b'^2) \nonumber \\ 
&&+ 2a^3a''b^2b' + 2a^3a'b^2b'' + 2a^4bb'b''  \;\;\;\;\;\;\;
\eea
and the equations of motion of scalar fields are given by

\bea
\label{gbdil}
\partial_r (a^2b^3 \partial_r \phi_i) & = & \frac{\partial_i V_{eff}}{b^3} -12 \partial_i G\;\bigg[ -a'^2b + 3a^2a'^2bb'^2 - a(ab''+2a'b'+a''b) \nonumber \\
     &&+ a^3b'\bigg( (ab''+a''b)b'+a'(2b'^2+2bb'') \bigg)\bigg]
\eea
\subsection{Zeroth order solution}

Consider constant scalar fields  $\phi_i=\phi_{i0}$. Then, equation ( \ref{eqgb1} ) can be solved by
\bea
\label{b}
b(r)=r
\eea
 Solving equations (\ref{eqgb1}) and (\ref{eqgb3}) for a double horizon (extremal) solution gives 
 (see the Appendix for details)
\bea
\label{a}
a^2(r) = 1+ \frac{r^2}{4 G_0} -\frac{r^2}{4G_0} \sqrt{1-\frac{8 G_0}{l^2} + \frac{16G_0 (G_0 + r_H^2 + \frac{3r_H^4}{2l^2})}{r^4} -\frac{8G_0r_H^4(1+\frac{2r_H^2}{l^2})}{r^6}}
\eea
where $G_0 = G(\phi_{0i})$. Notice that in this case it is easier to solve the Hamiltonian constraint 
(\ref{eqgb3}) than the equations of motion (as was done in section 2), since the equations of motion 
contain complicated second order derivative terms.  
Using the solutions for $a$ and $b$, the dilaton equation (\ref{gbdil}) becomes 
$\partial_i W (\phi_{i})|_{\phi_{i0}} =0$, where
\beq
W(\phi_i) = V_{eff}(\phi_i) + 12 r_H^4(1+\frac{3r_H^2}{l^2}) \ln \left(1+\frac{4G(\phi_i)}{r_H^2}\right)
\eeq
is the analogue of the "effective potential" when we add the Gauss-Bonnet correction.

\vspace{2mm}

The conditions for having attractor solutions  are
\bea
\label{W}
\partial_i W (\phi_{i})|_{\phi_{i0}} =0  
\eea

where
\beq
  \tilde{M}_{ij} = \frac{1}{2} \partial_i \partial_j W(\phi_{i0})\label{wmatrix}
\eeq
have positive eigenvalues.

\vspace{2mm}

To find the entropy, we write equation (\ref{eqgb3}) for the solution (\ref{b}) and (\ref{a}) at horizon $r=r_H$
and we get

\bea
\label{eqvef}
-1+ \frac{V_{eff}}{3 r_H^4} - \frac{2 r_H^2}{l^2} =0
\eea

Solving the algebraic equations (\ref{W}) and (\ref{eqvef}) together, gives $r_H$ and $\phi_{i0}$ in terms of 
the charges carried by black hole. The entropy is obtained from the entropy function, by adding
 $-2\pi \int_{S^3}{\cal L}_{GB}$ to it, for the metric in (\ref{adshor}).
\beq
S_{BH} = \frac{1}{2G_N} \pi^2 r_H^3(1+\frac{12 G_0}{r_H^2}) = 8 \pi^3 r_H^3(1+\frac{12 G_0}{r_H^2})
\eeq
\subsection{First order solution}
Starting with first order perturbation theory
\beq
\delta \phi_i = \phi_i -\phi_{i0} = \epsilon \phi_{i1}
\eeq
where $\epsilon $ is a small parameter we use to organize the perturbation theory.  The first correction to the scalars $\phi_i$ satisfies the equation
\beq
\partial_r (a^2 b^3 \partial_r \phi_{i1}) = \frac{ \beta_i^2}{ b_0^3} \phi_{i1}
\eeq
where $\beta_i^2$ is the eigenvalue for the matrix $2\tilde{M}_{ij}$. We are interested in a solution which does not blow up at the horizon $r=r_H$. This gives the following solution near the horizon
\beq
\phi_{i1} \simeq c_{1i} (1-\frac{r_H}{r})^{\gamma_i}
\eeq
where $\gamma_i$ are positive roots of 
\beq
\label{gamma2}
\gamma_i (\gamma_i+1) = \frac{\beta_i^2}{4 r_H^4} (1+\frac{4G_0}{r_H^2}) (1+ \frac{3r_H^2}{l^2})^{-1}
\eeq
Asymptotically (as $r\rightarrow \infty$) $\phi_{i1}$ takes a constant value, $c_{1i}$ --- however $\phi_{i1}$ is vanishing
at the horizon and the value of the scalar is fixed at $\phi_{i0}$ regardless of its value at infinity.
\subsection{Higher order solution}
\begin{figure}
 \begin{center}
  \includegraphics[scale=.6]{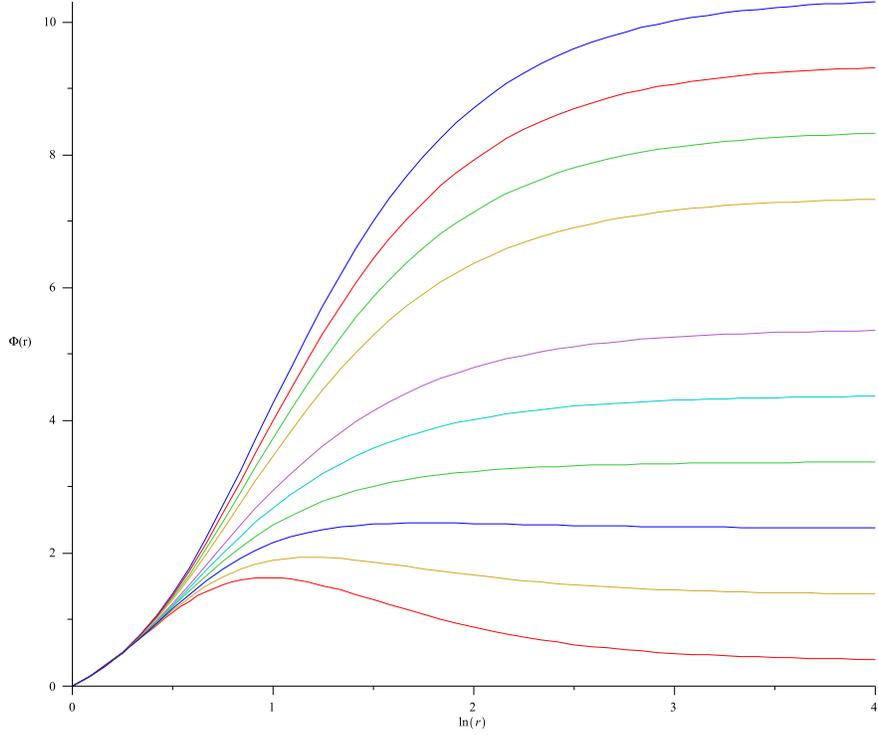}
 \end{center}
\caption{$\phi(r)$ vs. $\log(r)$, where numerical coefficients for the potentials are $g_2=1/9, v_2=2, v_0= 4$ and $g_0=1/4$. Different curves represent different asymptotic values for $\phi_{\infty}$. The attractor point is $\phi_0=0$ at the horizon,  $r_H=1$.}
\end{figure}
The analysis of the
 higher order solution is quite similar to the previous section. However it is rather difficult to solve the resulting differential equations analytically even in the second order. But as we will see below we can still solve our differential equations approximately order by order.

Without loss of generality, here we just consider the case with a single scalar field. All results can be simply generalized to the multi-scalar case. 
We can expand the solution in terms of the small parameter $x=r-r_H$ 
as a Frobenius series as follows
\bea\label{aexpan}
a(r)&=&\;(x+x^{\gamma_1}\;\sum_{n=2}^{\infty}a_n x^n),  \\
\label{bexpan}
b(r)&=&\; \frac{r_H}{1-x}\;(1+x^{\gamma_2}\sum_{n=1}^{\infty}
b_nx^n),  \\
\label{phiexpan}
\phi(r)&=&\; (\phi_0+x^{\gamma_3}\;\sum_{n=1}^{\infty}\phi_n x^n),
\eea  We also take a 
common $\gamma_i\equiv \gamma$ for all the solutions and write the series as: 
\bea
\phi(r)&=& \phi_0+K x^\gamma +\cdots,   \cr\cr
a(r)&=& x + a_1 x^{\gamma+1}+\cdots,  \cr\cr
b(r)&=& r (1+ b_1 x^\gamma+\cdots ). \nonumber
\eea
We also consider Taylor series expansions for $V_{eff}(\phi)$ and $G(\phi)$ as follows,
\bea \label{Vtaylor}
V_{eff}(\phi)&=&v_0 + v_1 (\phi-\phi_0)+ \frac{1}{2}v_2 (\phi-\phi_0)^2 +\cdots ,\\
\label{Gtaylor}
G(\phi)&=&g_0 + g_1 (\phi-\phi_0)+ \frac{1}{2}g_2 (\phi-\phi_0)^2 +\cdots \:.
\eea
By a careful investigation near the horizon, for the lowest power of $x$ which is $x^\gamma$, one can solve the set of equations as we did in the previous subsection and find the non-trivial solutions (from (\ref{gamma2}), (\ref{eqvef})
and (\ref{wmatrix}))
\bea \label{lambdaGB}
\gamma&=&\frac{1}{2}\left(-1+\sqrt{1+\frac{\beta^2}{r_H^4}(1+\frac{4g_0}{r_H^2})(1+\frac{3r_H^2}{l^2})^{-1}}\right) , 
\eea
\bea  \label{attractor2}
v_0=3r_H^4 + \frac{6 r_H^6}{l^2}, \;\;\;\;\;\; 
\eea
with $v_1=g_1=a_1=b_1=0$ and
\be 
\beta^2= v_2 + \frac{48 g_2 r_H^4(1+\frac{3r_H^2}{l^2})}{r_H^2+ 4 g_0}
\ee 

However, $v_2$, $g_2$ and $K$ are undetermined to this order.  The second equation in (\ref{attractor2}) 
is the extremum condition for $W$ which gives the attractor value $\phi_0$ at the horizon. Notice that we 
are faced with an extra condition $g_1=0$, which indicates that $G$, $V_{eff}$ and $W$ are at their 
extremum at the horizon, simultaneously. Such a case is
 the only situation where a non-integer $\gamma$ can be found. Otherwise we have to choose $\gamma=0$ for $g_1 \neq 0$. 
\vskip 0.2cm
The regularity condition for $\phi$ indicates that $\gamma$ should be non-negative and it in turn gives $\beta^2>0$, or
\be 
v_2 + \frac{48 g_2 r_H^4(1+\frac{3r_H^2}{l^2})}{r_H^2+ 4 g_0} \ge 0 \;.
\ee 
This again means that $W_H$ is minimum at its extremum point $\phi_0$.
\vskip 0.2cm
Higher order terms can be derived in a similar fashion. The important point is that, 
due to the non-linear nature of equations, they are a mixture of different powers of $\gamma$, like $x^{n\gamma}$ as well as $x^{n\gamma+m}$. To order these powers, we assume $0<\gamma <1$. Then the next leading term would be $x^{2\gamma}$. For higher order terms, since $\gamma$ is already known from the first order result (\ref{lambdaGB}), we can determine whether the next order is $x^{3\gamma}$ or
 $x^{\gamma+1}$. For small enough $\gamma$, it shows that we are generating a power series, $x^{n\gamma}$ as 
argued in \cite{Goldstein:2005hq}. 
\vskip 0.2cm
Notice that in contrast to the analysis of the 
previous section, here we considered all the equations simultaneously. This first means that, in principle, we are taking the backreaction into account. Secondly, we are dealing with a higher derivative theory, besides the  Klein-Gordon equation for the $\phi$ field. Other equations also involve the second derivative of $\phi$ and are important in the dynamics of $\phi$. So, 
they should be investigated as well.  To avoid quoting lengthy results, we demonstrate our results 
for a numerical simulation in figure 3.

\section{Embedding in string theory: attractor horizons and moduli `flow'}
In this section we present some physical interpretations of our results in the context of the AdS/CFT 
correpondence. After constructing the embedding in string theory, we consistently interpret the moduli 
flow in the bulk as a `holographic' RG flow and construct the $c$-function. The attractor horizon has 
spherical topology and corresponds to an IR fixed point. 

\subsection{Holographic RG flow}

We start by reviewing some known useful facts about the RG flow  --- we 
discuss the RG flow within the AdS/CFT correspondence and then 
we will interpret the moduli flow as a `holographic' RG flow in the bulk of AdS spacetime.

The RG equation of a system (represented by its initial set of coupling constants) describes a trajectory 
(`flow') in the coupling constants space. The set of all such trajectories generated by different initial sets 
of coupling constants define the RG flow in the coupling constants space. In general it is found that such a 
trajectory is attracted to a fixed point that is a functional attractor for the flow. The behaviour within the 
functional attractor is then determined by the $\beta$-function for the relevant couplings. In string theory 
the couplings are identified with the moduli space of the theory under consideration.\footnote{The constants which 
appear upon compactification are vacuum expectation values of certain massless fields. Thus, they are 
determined dynamically by the choice of the vacuum, i.e. the choice of the consistent string background.}

The AdS/CFT correspondence is referred to as a duality since
the supergravity (closed string) description of D-branes and the field
theory (open string) description are different formulations of the same
physics. This way, the infrared (IR) divergences of quantum gravity in the
bulk are equivalent to ultraviolet (UV) divergences of the dual field theory
living on the boundary. A remarkable property of the AdS/CFT correspondence is that it 
works even far from the conformal regime. 
Conformal field theories in various dimensions correspond to $AdS_{d+1}\times X_q$ gravitational theories.
But one can also have cases that interpolate between asymptotically AdS spaces at the boundary and in the middle 
of the bulk, that are naturally interpreted as two conformal points of a dual QFT.
Any hypersurface of constant radius in 
the bulk of AdS should have a field theory dual and the radial coordinate is consistently 
interpreted as the energy scale in the field theory. The RG `trajectory' then allows us to
define the UV and the IR limits of a given QFT and in the dual to interpret the `radial' flow as a holographic 
RG flow. At a critical point a system can be regarded as scale invariant due to the violent critical
 fluctuations of the order parameter which lack any characteristic length and time scale. Thus, in the 
AdS/CFT context, the CFT on the boundary is the UV fixed point of a QFT in the bulk. Using the gravity 
side of the correspondence (deformations of AdS) one can obtain holographic RG flows corresponding to 
non-conformal field theories. 

In \cite{de Boer:1999xf}, the Hamilton-Jacobi equations of canonical gravity were used to 
obtain first-order differential equations\footnote{In the context of attractor mechanism in flat spacetime, 
attempts to obtain first-order differential equations and interpret non-BPS extremal black hole solutions 
were made in \cite{Miller:2006ay, Ceresole:2007wx}.} from the supergravity equations of motion and to derive the 
holographic RG flow. Specifically, the authors of \cite{de Boer:1999xf} studied an action of scalar fields 
coupled to gravity with a non-trivial potential for the scalars:
\be
S_{\rm loc}\, [\, \phi\, ,\, g\, ] \, =  \;
\int \!\! \sqrt{g}  \, \Bigl(V(\phi)  +   
R \, +  \frac{1}{2}\partial^\mu\phi^i g_{ij}(\phi)\, 
\partial_\mu \phi^j \Bigr)
\ee
The first order equations of motion can be written as 
\be 
\dot{\phi}^i = g^{ij} \partial_j U , \qquad 
\dot{g}_{\mu\nu} = -\frac{1}{3}U(\phi) g_{\mu\nu} 
\ee
and the scalar potential $V(\phi)$ is related to `superpotential' $U(\phi)$ by
\be
\label{fst}
V =  \frac{1}{3}
U^2 - \frac{1}{2}\partial_i U \, g^{ij} \partial_j U
\ee
The identification of the holographic RG flow with the field theory local RG flow expressing how Weyl 
symmetry is broken allows to construct a holographic $c$-function and also the $\beta$-function. 
The solutions of this theory are BPS domain walls (${\cal N}=1$ supersymmetric kinks in the radial direction)
and the flow is between different AdS spaces, which correspond to different ground states of the 5 dimensional 
${\cal N}=8$ gauged supergravity. 

However, the attractor mechanism appears in a slightly different context: the moduli potential 
is trivial (a constant) and instead
the gauge fields in the bulk are turned on. There is an induced effective potential for 
the moduli due to the non-trivial coupling between moduli and the form fieds. The non-BPS flow is now 
between the boundary of an $AdS_5$ (UV region) and the horizon of the 
extremal black hole (IR region). For the extremal black hole solution, we have the usual $AdS_5$, but a 
trivial flow. For a large enough perturbation $\delta \phi_i$, we can reach another $AdS_5$ vacuum, and then 
we will have a holographic flow between the new $AdS_5$ and the horizon of the extremal black hole. 

There is an enhanced 
symmetry $AdS_2\times S^3$ in the near-horizon limit and so the flow is still between two AdS vacua, albeit 
with different dimensionality --- 
however,the supersymmetry can be broken in the bulk. The breaking of supersymmetry is not problematic, 
as one can have a nonsupersymmetric RG flow between conformal points in a supersymmetric theory, but 
the change from $AdS_5$ to $AdS_2\times S^3$ is more puzzling, however we will analyze it in the next subsection.

At this point it is useful, though, to present some computational details on solutions which interpolate between two 
critical points and emphasize similarities between the kink solutions with and without horizons.

First, let us discuss a simple model with a single bulk scalar field coupled to gravity (see e.g., \cite{fgpw,df}). 
We describe a domain wall solution that interpolates between two AdS spaces with different radii.
The 
scalar non-linear equation of motion is well approximated by a linear solution near a critical point --- let us consider 
a quadratic approximation given by:
\beq
V(\phi)\simeq V(\phi_0)+\frac{1}{2}\beta^2\phi_1^2\, , \hspace {2cm} \phi_1=\phi-\phi_0
\eeq
thus the corresponding solution is (here $r$ is a coordinate in the AdS space at infinity 
defined as in (\ref{meatin}))
\beq
\phi_1(r)=Ar^{(\Delta-4)}+Br^{-\Delta }\, , \hspace {2cm} \Delta=2+\sqrt{4+\beta^2l^2}
\eeq
The UV point corresponds to the boundary ($r\rightarrow \infty$) and the fluctuation should die off
($\phi(r)\rightarrow \phi_0^{UV}$). The solution is then given by 
\beq
\phi(r)=\phi_0^{UV}+\phi_1^{UV}\approx \phi_0^{UV}+Ar^{(\Delta^{UV}-4)}+Br^{-\Delta^{UV}} 
\eeq
with the constraint $2<\Delta^{UV}<4$ that is equivalent with a negative mass$^2$, $\beta^2<0$.\footnote{Due to 
the negative curvature, fields with negative mass are permitted to exist in AdS. In fact  the lower bound 
$\beta^2>-4$ corresponds to the stability bound for field theory in Lorentzian AdS.} 
Thus the critical point is a local maximum given by $V(\phi_0^{UV})$ and the dual QFT 
is a {\it relevant} deformation of an UV CFT that is living on the boundary region of the domain wall solution. 

The IR point corresponds to a region deep in the bulk ($r\rightarrow 0$) and the corresponding critical 
point should be a minimum. The solution is again given by 
\beq
\phi(r)=\phi_0^{IR}+\phi_1^{IR}\approx \phi_0^{IR}+Cr^{(\Delta^{IR}-4)}+Dr^{-\Delta^{IR}} 
\eeq
except that $\Delta^{IR}>4$ that correponds to a positive mass$^2$, $\beta^2>0$ --- that imposes a further 
constraint, namely $D=0$ (this term would be divergent). Thus, as expected from RG flow properties, the domain 
wall approaches the IR region in the bulk with the scaling rate of an {\it irrelevant} operator of  
dimension $\Delta^{IR}>4$. 

We are now ready to understand the case of interest, a black hole in AdS. In this case the deep IR 
region corresponds to the black hole horizon. By imposing the attractor conditions, the horizon should be 
a stable minimum of the effective potential. It is interesting to observe that there also are two solutions 
in the near horizon limit, but the existence of a regular horizon forces us to discard the divergent mode.
Therefore, the attractor horizon describes the IR point of an RG flow and corresponds to a deformation of a 
CFT by an {\it irrelevant} operator (see, also, \cite{Goldstein:2005hq}). For completness, we present 
the behaviour of the first order solution at the AdS boundary. Consider the equations (\ref{a0}) and (\ref{phi2}) at 
large $r$ ---  (\ref{phi2}) becomes 
\beq
\partial_{r} ( \frac{r^5}{l^2} \partial_{r} \phi_{i1} ) =  \frac{\beta_{i}^2}{r^3} \phi_{i1}
\eeq
Let us define $y \equiv \frac{\beta_{i} l}{3 r^3}$, then for large r we obtain 
\beq
\phi_{i1}(r)=c_{1} y^{\frac{2}{3}} I_{\frac{2}{3}}(y) + c_{2} y^{\frac{2}{3}} I_{-\frac{2}{3}}(y)
\eeq
where $c_{1}$ and $c_{2}$ are arbitrary constants and $I_{\nu}$ stands for a modified Bessel function.

In conclusion it is very tempting to interpret the moduli flow as the holographic RG flow in the bulk --- we will make 
these ideas more concrete in the following by constructing explicitly the string theory embedding of our system 
and studying its $c$-function. 

\subsection{String theory embedding}
We have been analyzing asymptotically $AdS_5$ solutions until now.
In order to talk about AdS/CFT however we need to have 10 dimensional IIB supergravity solutions.  So we need to 
understand whether we can embed the extremal black holes in 10 dimensions via a consistent truncation. The 
extremal black holes can be embedded in 5 dimensional ${\cal N}=8$ gauged supergravity, and therefore are in the 
same class as the solutions of \cite{fgpw}. It is believed that the full ${\cal N}=8$ 5 dimensional gauged
supergravity is a consistent truncation of 10 dimensional IIB supergravity
as in the 4 dimensional \cite{dn} and 7 dimensional \cite{nvv} cases, though until now only subsets of it 
have been obtained as consistent truncations. For the case of extremal black holes however, we have not only an
embedding in 10 dimensional IIB supergravity, but we can even obtain it as the near horizon limit of a system 
of D-branes \cite{Duff:1999rk}.

The extremal RNAdS solution is a special case of the 3-charge black holes in \cite{Behrndt:1998jd, Duff:1999rk}, with $H_1=H_2=H_3\equiv H$.
The general (non-extremal) solution is 
\bea
&&ds_5^2=-(H_1H_2H_3)^{-2/3}f\; dt^2+(H_1H_2H_3)^{1/3} (f^{-1}dr^2+r^2 d\Omega_{3,k}^2);
\nonumber\\
&&X_i=H_i^{-1}(H_1H_2H_3)^{1/3},\;\;\; A^i=\sqrt{k}(1-H_i^{-1})\coth \beta_i dt
\eea
where
\be
f=k-\frac{\mu}{r^2}+g^2r^2(H_1H_2H_3),\;\;\; H_i=1+\frac{\mu\sinh^2\beta_i}{kr^2}
\ee
and $k=1,0,-1$ corresponds to having $S^3,T^3$ or $H^3$ foliations, so the case studied here corresponds to $k=1$. 
If $H_1=H_2=H_3=H$ (thus $\beta_i=\beta$), then $X_i$=constant=$X$.
The change of coordinates $\tilde{r}^2=Hr^2=r^2+\mu\sinh^2\beta$ brings us to our metric in $\tilde{r}$ coordinates,
since $dr^2=H d\tilde{r}^2$ and 
\be
H^{-2}f=g^2\tilde{r}^2+1-\frac{\mu(2\sinh^2\beta_i+1)}{\tilde{r}^2}+
\frac{\mu^2\sinh^2\beta_i\cosh^2\beta_i}{\tilde{r}^4}
\ee
is identified with $a^2$, if we have
\be
g^2=\frac{1}{l^2};\;\;
\sinh^2\beta=-\frac{1}{2}+\frac{1}{2}\sqrt{1+\frac{1+2r_H^2/l^2}{r_H^2/l^2(1+9 r_H^2/4l^2)}};\;\;\;
\mu=\frac{r_H^2(2+3r_H^2/l^2)}{\sqrt{1+\frac{1+2r_H^2/l^2}{r_H^2/l^2(1+9 r_H^2/4l^2)}}}
\ee
and then we also get $A_i=q dt/\tilde{r}^2$.

The 10 dimensional embedding of the extremal RNAdS solution is obtained from the 10 dimensional reduction ansatz 
used in \cite{Duff:1999rk}, namely
\be
ds_{10}^2=\sqrt{\tilde{\Delta}}ds_5^2+\frac{1}{g^2\sqrt{\tilde{\Delta}}}\sum_{i=1}^3X_i^{-1}
(d\mu_i^2+\mu_i^2(d\phi_i+gA_i)^2)\label{10dans}
\ee
where $\tilde{\Delta}=\sum_{i=1}^3X_i\mu_i^2$ and
\be
d\Omega_5^2=\sum_{i=1}^3d\mu_i^2+\mu_i^2d\phi_i^2;\;\;\;
\mu_1=\sin\theta;\;\;\mu_2=\cos\theta\sin\psi ;\;\;\;
\mu_3=\cos\theta\cos\psi
\ee
is the 5-sphere metric. It is important to notice that each angular momentum becomes a charge after KK reduction 
on $S^5$ --- this resembles the KK reduction on a circle when the momentum on the circle becomes the electric 
charge (the circle fibration will give the magnetic charge).

In our case, since $X_i=X$=1, $\tilde{\Delta}=X$, and $1/g^2=l^2$ we get
\be
ds_{10}^2=ds_5^2+ l^2\sum_{i=1}^3[d\mu_i^2+\mu_i^2(d\phi_i+g A_i)^2]
\ee
so the extremal RNAdS is embedded (up to a constant rescaling) by just adding a sphere of radius $l$, squashed by 
the gauge field, i.e. rotating on this 5-sphere.

Since we have a 10 dimensional IIB supergravity solution, we can safely use AdS/CFT. But it would be useful to 
have also a D-brane solution that gives the above solution in the  decoupling limit.

It is known that the general
RNAdS black hole embeds to the above 10 dimensional metric, which corresponds to adding 
a chemical potential for the R charge in AdS/CFT \cite{Hawking:1999dp} . It is also known how to 
obtain the $k=0$ (torus foliation) AdS black holes from the decoupling limit of rotating D3-branes 
\cite{Duff:1999rk}. But a 
minimal change is needed to obtain the $k=1$ black holes considered here. 

The D3 branes rotating with 3 angular momenta $l_i,i=1,2,3$ in 3 different directions have the metric
\bea
&&ds^2=H^{-1/2}(-(1-\frac{2m}{r^4\Delta})dt^2+dx_1^2+dx_2^2+dx_3^2)+H^{1/2}\left[
\frac{\Delta dr^2}{H_1H_2H_3-2m/r^4}\right.\nonumber\\
&&\left. +r^2\sum_{i=1}^3H_i(d\mu_i^2+\mu_i^2d\phi_i^2)
-\frac{4m\cosh \alpha}{
r^4H\Delta}dt \sum_{i=1}^3 l_i \mu_i^2d\phi_i+ \frac{2m}{r^4H \Delta}(\sum_{i=1}^3
l_i\mu_i^2d\phi_i)^2 
\right]\label{rotd3}
\eea
where
\be
\Delta=H_1H_2H_3\sum_{i=1}^3\frac{\mu_i^2}{H_i};\;\;\;
H=1+\frac{2m\sinh^2\alpha}{r^4\Delta};\;\;\;
H_i=1+\frac{l_i^2}{r^2}
\ee
In our case, $l_1=l_2=l_3\equiv l_0$, so $H_1=H_2=H_3\equiv h$, and $\Delta = h^2$. 
Making again the change of variables $\tilde{r}^2=r^2+l_0^2$ we get $r^4\Delta=\tilde{r}^4$ 
and
\bea
&&ds^2=H^{-1/2}[-(1-\frac{2m}{\tilde{r}^4})dt^2+dx_1^2+dx_2^2+dx_3^2]+H^{1/2}\left[
\frac{d\tilde{r}^2}{1-\frac{2m}{\tilde{r}^4}+\frac{2ml_0^2}{\tilde{r}^6}}\right.
\nonumber\\&&\left.
+\tilde{r}^2d\Omega_5^2-\frac{4m l_0\cosh\alpha}{\tilde{r}^4H}dt\sum_{i=1}^3\mu_i^2
d\phi_i+\frac{2m l_0^2}{\tilde{r}^4H}(\sum_{i=1}^3\mu_i^2d\phi_i)^2\right]
\eea
where $H=1+2m \sinh^2\alpha \; \tilde{r}^{-4}$.

The decoupling limit is obtained via the rescalings
\be
m=\epsilon^4 m';\;\;\;
\sinh\alpha =\epsilon^{-2}\sinh \alpha ';\;\;\;
r=\epsilon\; r';\;\;\; x^{\mu}=\epsilon^{-1} x'^{\mu};\;\;\;
l_i=\epsilon l_i '\label{decoupling}
\ee
followed by $\epsilon\rightarrow 0$ and dropping the primes. One then gets
(\ref{10dans}) with 
\be
d\Omega_{3,k}^2=d\vec{y}\cdot d\vec{y};\;\;\;
\vec{y}=g\vec{x};\;\;\;
\frac{1}{g^2}=\sqrt{2m}\sinh \alpha ;\;\;\;
\mu=2m g^2;\;\;\; l_i^2=\mu\sinh^2\beta_i
\ee

But notice that if in the final metric (\ref{10dans}) we change from $k=0$ to $k=1$, 
replacing $d\vec{y}\cdot d\vec{y}$ with $d\Omega_3^2$, and correspondingly we add 
$1$ inside $f$, we can still obtain the metric from the decoupling limit of the same 
D3 brane metric  (\ref{rotd3}), with an infinitesimal perturbation. 
Indeed, the decoupling limit involves $x^{\mu}=\epsilon^{-1}x'^{\mu}$, which now we 
can understand as rescaling the radius in $R^2d\Omega_3^2$ by $R=R'/\epsilon$. 
Since after the rescaling we want to have $R'=1$, it means initially $R=1/\epsilon
\rightarrow \infty$, thus a sphere of a very large radius, still approximated by a 
plane. Moreover, the addition of the $+1$ in $f$ implies the addition of 
$(g^2r^2 H_1H_2H_3)^{-1}$ to $1-\frac{2m}{r^4\Delta}$ in the coefficient of $dt^2$
and of $1/(g^2r^2)$ ($=\sqrt{2m}\sinh \alpha/r^2$)
to $ H_1H_2H_3-2m r^{-4}$ in the coefficient of $dr^2$. However,
the same reasoning tells us that in order to get a finite result after the decoupling 
limit, these added terms need to be multiplied by $\epsilon^2\rightarrow 0$.
In conclusion, we can actually also obtain the $k=1$ case (sphere foliation) from 
(\ref{rotd3}), with an infinitesimal perturbation. 

So the extremal RNAdS metric with sphere foliation can be obtained as a consistent 
truncation of the near horizon limit of rotating D3 branes. Since the nonrotating 
case is dual to ${\cal N}=4$ Super Yang-Mills, the perturbed extremal RNAdS solution should 
correspond to adding a perturbation in the UV  and 
obtaining an RG flow to a different (IR) conformal fixed point. 

As we noticed however, there is an aparent discrepancy in dimensionality, namely 
we start with $AdS_5$ in the UV and get $AdS_2\times S^3$ in the IR, that seems 
problematic at first sight (for usual holographic flows we go between two different $AdS_5$'s).
 However, what is important for the conformal field 
theory is the global boundary of $AdS_5$ and $AdS_2\times S^3$, which in both 
cases is $R_t\times S^3$ (conformal to 4 dimensional flat space). 
Note that we have used throughout the sphere foliation of AdS space and the AdS space extremal black hole, where 
the metric at infinity is 
\be
ds^2\simeq -\frac{r^2}{l^2}dt^2+l^2\frac{dr^2}{r^2}+r^2d\Omega_3^2\label{meatin}
\ee
for which the $R_t\times S^3$ boundary at $r\rightarrow \infty$ is actually parametrized by $t$ and $\Omega_3$
(and $r^2$ is taken out when we consider the conformal boundary).
For the $AdS_2\times S^3$ near the horizon, the $S^3$ metric is $(r_H+\rho)^2d\Omega_3^2$ (here $\rho=r-r_H$), 
and the interpolation between $r^2$ and $r_H^2$ is done by $b(r)^2$, which does look indeed like a holographic flow 
between two dual RG fixed points. Notice then that the correct conformal radius of the two (dual) $\Omega_3$'s is 
given by $b'(r)$, since it gives $1$ for the extremal black hole, both at the boundary and at infinity. This is 
what we want since in the extremal case the dual CFT does not change, and has conformal radius $1$.

In conclusion, in the case of a nontrivial holographic flow (perturbed extremal black hole) both 
the UV and IR correspond to 4 dimensional conformal field theories, which we expected,
since we have a holographic flow of ${\cal N}=8$ 5 dimensional gauged supergravity, 
dual to an RG flow in 4 dimensions. A nontrivial holographic flow occurs if the perturbation $\delta\phi_i$ is 
large enough to produce a new $AdS_5$ vacuum. This is so, since an RG flow relates two conformal fixed points, 
and a small perturbation will get us away from the original $AdS_5$ vacuum.

The holographic RG flow is 10 dimensional, but reduces to the 5 dimensional flow 
upon the dimensional reduction on $S^5$ in (\ref{10dans}). The sphere $S^5$ now plays a 
role, since its squashing (rotation) due to the gauge field $A_i$ is partly responsible
for the flow (unlike previous cases, the flow is not solely governed by scalar fields, but is 
also governed by  gauge fields). 

Another important consequence of the embedding of the 5 dimensional extremal RNAdS 
solution into a rotating D3 solution is understanding the attractor mechanism from 
a different perspective. The 10 dimensional rotating D3 branes will presumably 
have their own attractor mechanism, i.e. a flow between the 10 dimensional 
flat space at infinity and the near horizon limit of the rotating D3 branes. 
One could conceivably then embed the 5 dimensional AdS space attractor mechanism 
discussed in this paper as the decoupling limit (\ref{decoupling})
of the attractor mechanism of the 
rotating D3 branes. We leave however the exploration of this possibility for further
work. 

\subsection{c-function}

We now turn to the calculation of the c-function. A c-function is a monotonic function 
that takes the value of the central charge of the UV fixed point in the UV and 
of the IR fixed point in the IR.

The central charge counts the number of massless degrees of freedom in the CFT (it counts the ways in which 
the energy can be transmitted). The coarse graining of a quantum field theory removes the information about the 
small scales, in other words there is a gradual loss of non-scale invariant degrees of freedom. Thus, for a QFT RG flow, there should 
exist a c-function that is  decreasing monotonically
from the UV regime (large radii in the dual AdS space) where it gives $c_{UV}$ to the IR regime (small 
radii in the gravity dual) of the QFT where it gives $c_{IR}$, a statement known as the c-theorem. 
A c-theorem for gauge theory that is living on the AdS boundary with topology 
$R\times S^3$ was constructed in \cite{Elvang:2007ba}.\footnote{A c-function for charged (multi-)black holes in dS space 
was presented in \cite{Astefanesei:2003gw}.} 

In order to get the c-function we look for a monotonic function of $r$, $A(r)$, along the flow, and  then the c-function
is ${\mathcal C}(r)=A(r)^n$, such that 
\be
\frac{c_{UV}}{c_{IR}}=\frac{{\mathcal C}(r=\infty)}{{\mathcal C}(r=r_H)}=\left(\frac{A(r=\infty)}{A(r=r_H)}\right)^n
\ee
from which we find the appropriate power of $n$. 

The monotonic function of $r$ along the flow is found from the Einstein equations. The right hand side of the Einstein 
equations is the energy momentum tensor, which should obey the weak (null)
 energy condition, $T_{\mu\nu}\xi^{\mu}\xi^{\nu}\geq 
0$, with $\xi $ null. The weak energy condition (the GR equivalent of the positivity of local energy density) 
implies a second order inequality for the metric coefficients, that may sometimes be written as the positivity of 
the derivative of a function, thus extracting the monotonic function.

Consider the most general ansatz for the metric with spherical symmetry as follows
\be 
ds^2 = -a(r)^2 dt^2 + \frac{dr^2}{c(r)^2} +b(r) d\Omega_3^2
\ee 
The combination $R_{rr} -\frac{G_{rr}}{G_{tt}} R_{tt}$ of the Ricci tensor components gives
\be 
\label{Cf}
3\left(-\frac{b''}{b} - \frac{b'c'}{bc} + \frac{a'b'}{ab} \right) =g_{ij}\partial_{\mu}\phi^i\partial^{\mu}\phi^j =  \phi'^2
\ee 
using (\ref{Cf}), it is straightforward to see that
\be 
{\mathcal C}(r) = {\mathcal C}_0 \; \frac{a^3}{b'^3 c^3}
\ee 
is a monotonically increasing function of $r$ for any positive constant ${\mathcal C}_0$. Therefore for the 
unperturbed extremal solution ($a=c,b'=1$), ${\mathcal C}(r)={\mathcal C}_0=$ constant, which is as it should be, 
since for the extremal solution there is no RG flow. For the perturbed extremal black hole, $a=c$, but $b'\neq 1$, 
thus we get a nontrivial flow. The flow relates two different conformal fixed points if the perturbation is large 
enough to reach a new $AdS_5$ vacuum of ${\cal N}=8$ supergravity.
Since $b'(r)$ acts as the conformal radius of $S^3$, $c_{UV}/c_{IR}$ should 
indeed scale as $b'^3$, hence the cubic power in ${\mathcal C}(r)$.

\section{Discussion}
In this paper, we have investigated the attractor mechanism in spaces with 
negative cosmological constant. A straightforward extension of the effective 
potential method \cite{Goldstein:2005hq} confirms that the attractor mechanism 
still occurs for 5-dimensional extremal black holes in AdS space. This is expected 
since the near-horizon geometry of five dimensional extremal charged black holes 
in AdS has the $SO(2,1)\times SO(4)$ isometry. The origin of the attractor mechanism 
is in the enhanced symmetry of the near-horizon geometry that contains the infinite 
long throat of $AdS_2$. The entropy function is constructed 
(on an $SO(2,1)\times SO(4)$ symmetric background) by taking the Legendre transform 
(with respect to electric charges) of the reduced Lagrangian evaluated at the horizon. By 
extremising the entropy function one obtains the equations of motion at the horizon and 
its extremal value corresponds to the entropy that is independent of the asymptotic data. 
However, if the entropy function has flat directions the extremum remains fixed but the flat 
directions will not be fixed by near horizon data and can depend on the asymptotic moduli. 
In this paper we also have shown the equivalence of the effective potential method and 
the entropy function in the near-horizon limit for extremal black holes in AdS. 

In Section 3 we have studied the attractor mechanism in AdS space in the presence of higher 
derivative terms (we present explicit results for the Gauss-Bonet term). The analysis is more 
involved but we reached similar conclusions. The near-horizon geometry remains 
$AdS_2\times S^3$ even after adding $\alpha'$ corrections --- the radii of $AdS_2$ and $S^3$ 
receive corrections, but the geometry does not change (also see, e.g., \cite{Sen:2005wa, 
Sen:2005iz}).  For the Gauss-Bonnet correction, in 
a background with  asymptotic $AdS$ boundary condition, 
the regularity of scalar fields at the horizon\footnote{In fact the metric components should be 
analitic ($C^{\infty}$) in order to obtain a smooth event horizon. We used this condition in 
our numerical analysis.} is a sufficient (and obviously necessary) condition to have the attractor 
mechanism --- regularity at the horizon restricts the effective potential $W$ to be at its minimum 
at the critical point that is equivalent with the fact that the near horizon geometry is $AdS_2\times S^3$.

Sen's entropy function formalism was 
applied to extremal black holes in AdS in the presence of higher derivatives terms in 
\cite{Morales:2006gm}. The advantage of this method is that the higher derivatives terms can be 
incorporated easily, but the method can not be used to determine the properties of the solution 
away from the horizon. In this paper, we used the effective potential method (that is based on the 
equations of motion in the bulk) to prove the existence of the attractor mechanism in AdS space 
with higher derivatives. 

When the scalar potential is not a constant, a general analysis of the attractor mechanism is difficult.
The reason is that the right hand side of the equation of motion for the moduli (\ref{phi}) contains 
two terms: a term that depends of the effective potential and the other one depends on the scalar potential.
Thus, there is a competition between these two terms in the bulk and this is why the analysis is difficult.
In the near horizon limit both terms are present and if the near horizon geometry is still 
$AdS_2 \times S^3$, then the entropy function formalsim can still be applied to compute the entropy. On 
the other hand the effective potential dies off at the boundary and the moduli at the bounday are fixed 
at the minimum of the scalar potential --- the existence of a full solution from the horizon to the 
boundary is problematic in this case.  However, within the AdS/CFT correspondence, the critical point of 
the potential at the boundary should be a a local maximum such that a relevant deformation in the 
ultraviolet CFT gives a new long distance 
realization of the field theory. Therefore a discussion of the attractor mechanism for a theory with a 
non-trivial moduli potential should be made case by case. On the other hand, we were able to completely 
study the attractor mechanism in AdS space for which the moduli potential is constant (related to 
the cosmological constant). 

In Section 4 we provided some physical interpretations of our results in the context of the AdS/CFT correspondence. 
We interpreted the moduli flow as a holographic RG flow in the AdS bulk and constructed the corresponding c-function. 
Let us now discuss in detail these results. 

It is well known that by using different foliations of AdS space one can describe boundaries that have 
different topologies affording the study of CFT on different backgrounds. We are interested in a foliation 
of AdS for which the boundary has the topology $R\times S^3$ --- the black holes in this space have 
horizons with spherical topology. The diffeomorphisms in the bulk are equivalent with the conformal 
transformations in the boundary and the spherical boundary is related conformally to the flat boundary 
if the point at the infinity is added for the latter. In other words, the boundaries with different topologies 
are related by {\it singular} conformal transformations. Since the CFT is living on a boundary with 
the topology $R\times S^3$ there is an additional Casimir-type contribution to the total energy in accord 
with the expectations from quantum field theory in curved space \cite{Balasubramanian:1999re} --- the 
central charge is related 
to the Casimir energy in the boundary. In \cite{Elvang:2007ba}, a c-function (an off-shell generalization 
of the central charge) for an AdS space foliated by spherical slices was proposed. 

For a supersymmetric 
flow the IR point is a naked singularity (the BPS limit of a RN-AdS black hole is different 
than its extremal limit and it has a naked singularity) --- the analysis in \cite{Elvang:2007ba} was done for 
only one gauge field. There is a similar 
situation in flat space: if we consider a theory with just {\it one} gauge field exponentially coupled to the dilaton, the 
extremal limit is a naked singularity. The non-trivial form field generates an effective potential for the dilaton, but 
this potential does not have a minimum. To obtain a stable minimum a second gauge field should be turned on and 
so in theories with more than one gauge field we expect non-singular BPS limits of extremal black holes in AdS.
With just one gauge field, as in \cite{Elvang:2007ba}, we expect that the $\alpha'$ corrections will play an 
important role and the naked singularity may be dressed by a horizon. Another way to avoid this problem may
be that the flow is ending on the surface of a star\footnote{This is a  stable state without horizon, so at zero 
temperature --- AdS stars were constructed  in \cite{Astefanesei:2003rw}.} and 
so the number of degrees of freedom is much smaller than in the case of a horizon, but still non-zero.

However, our case is different --- we are interested in non-supersymmetric attractors in AdS for which 
the supersymmetry can be broken in the bulk. At a first look, the existence of a c-function seems problematic. This 
issue was addressed in \cite{Goldstein:2005rr} for attractor horizons in flat space (though, the physical interpretation 
of a c-function in flat space is less clear than in AdS) and is closely related with the existence of first order 
equations of motion. In \cite{Goldstein:2005rr} it was argued that the boundary conditions play an important role in 
the sense that the attractor boundary conditions restrict the allowed initial conditions to make the equations 
first order such that the solution at the horizon is regular. The definition of the c-function is best understood 
if we have a concrete string theory embedding.

We have shown that the extremal black hole becomes a 10 dimensional black hole solution rotating in the extra 
$S^5$ (we simply add a 5-sphere deformed by the gauge field). Also, we have seen that we can obtain this 
10 dimensional metric as the decoupling limit of a system of rotating black 
D3 branes, even in the case we are interested 
in, of boundary topology $R\times S^3$ (sphere foliation of the AdS black hole). Now, we can safely interpret 
our results within AdS/CFT duality as in the following.

Since the AdS black hole is the dimensional reduction of the decoupling limit of a black D3-brane system, which has 
a flat asymptotic at 10 dimensional 
infinity, one could also try to understand the attractor mechanism by embedding in flat 10 dimensions.
The rotating black D3-branes have a near horizon limit, which is however harder to understand because of the angular 
dependence. Therefore we can imagine defining an attractor mechanism for the rotating black D3-branes in 
flat 10 dimensional space, as in \cite{Astefanesei:2006dd}. If the 10d horizon is an attractor, one could 
imagine taking a decoupling limit 
of the near horizon geometry, followed by dimensional reduction, and hopefully one would obtain the 5 dimensional 
attractor (horizon of the 5d AdS black hole), thus embedding the $AdS_5$ attractor mechanism in the usual attractor 
mechanism in flat 10 dimensions. The limits involved are however quite subtle, so we leave the details of this 
analysis for further work.

To define AdS/CFT, one looks at fluctuations in the gravity dual. A field in Lorentzian AdS space
has two kinds of modes, normalizable and non-normalizabile. While the former  corresponds to 
a state in the CFT, the non-normalizable mode corresponds to insertion of an operator in the 
boundary (a bulk field is the source for an operator in the QFT). So, if the 
boundary conditions are kept fixed and the bulk is modified (for example, black holes or 
gravitational waves) the objects in the bulk correspond to states in the boundary (certain operators 
acquire expectations values). An extremal black hole is a zero temperature state in the CFT. An 
operator deformation in the 
CFT will produce an interpolating flow in which the scalar moduli approach a maximum critical 
point at the (UV) boundary and a minimum at the (IR) black hole horizon. 

We have shown that the attractor mechanism works also in AdS space, not only in flat space. 
However, the interpretation seems to be somewhat different. In flat space, the attractor mechanism 
means that the horizon values for the moduli of the extremal black hole are fixed and the entropy 
(the value of the entropy function at the minimum) 
depends only on the charges $q_i$. In $AdS_5$, a priori there is one more parameter in the entropy, 
$G_{N,5}/l^3$ ($l$ is the AdS radius), which varies continuosly. However, the correct interpretation of 
the moduli flow is as a 10 dimensional RG flow. Then, in string theory, $G_{N,5}=G_{N}/(R_{S^5})^5$ and 
$R_{S^5}=l, G_{N}=g_s^2(\alpha ')^4$, which means that $G_{N,5}/l^3=1/N^2$ is independent on any continous 
parameters. Therefore only after embedding in string theory the attractor mechanism is on the same footing
as in flat space.

\acknowledgments 
D.A. would like to thank Nabamita Banerjee, Suvankar Dutta, Rajesh Gopakumar, Eugen Radu,
Ashoke Sen, and Sandip Trivedi for interesting conversations and collaboration on related projects.
H.Y and S.Y. would like to thank Sangmin Lee, Soo-Jong Rey and Ashoke Sen  for discussions.  
The research of D.A. and H.N. has been done with partial support from MEXT's program
"Promotion of Environmental Improvement for Independence of Young Researchers"
under the Special Coordination Funds for Promoting Science and Technology. D.A. also 
acknowledges support from NSERC of Canada. The work of H.Y. and S.Y. is supported by the 
Korea Research Foundation Leading Scientist Grant (R02-2004-000-10150-0) and Star Faculty Grant (KRF-2005-084-C00003).

\appendix

\section{The equations of motion for the $R^2$ term}
\label{A1}
For 5-dimensional, spherically symmetric, extremal Gauss-Bonnet black holes with a cosmological constant, we take the
following ansatz for the metric:
\beq
  ds^{2}  = -a(r)^{2}dt^{2}+c(r)^{-2}dr^{2}+b(r)^{2}d\Omega_{3}^{2}
\eeq
Then all the equations of motion including the Hamiltonian constraint are obtained by taking derivatives with respect to the metric components, gauge and scalar fields and choosing the gauge $a(r)=c(r)$ from the following one-dimensional action:
\beq
I = \frac{\pi}{8 G_{N}} \int dr \frac{a b^3}{c}
   \bigg[R - c^2 g_{ij}\partial_r\phi^i\partial_r\phi^j -\frac{2}{b^{6}}V_{eff}(\phi^i) + \frac{12}{l^2} + G(\phi^i)L_{GB}\bigg]
\eeq
where
\beq
R = -\frac{2}{ab^2} \bigg[ b^2c(a'c)' + 3 a(-1 + c^2 b'^2) + 3 bc (ab'c)' \bigg]
\eeq
\beq
V_{eff}(\phi^i)=f^{AB}(\phi^i) Q_A Q_B
\eeq
and
\bea
L_{GB}=&&R^2-4R_{\mu\nu}R^{\mu\nu}+R_{\mu\nu\alpha\beta}R^{\mu\nu\alpha\beta} \nonumber \\
      =&&\frac{24c}{ab^3} \bigg[ -(ab)'c' + (ab'+3a'b)b'^2c^2c' - (ab''+a'b'+a''b)c \nonumber \\
      &&+ \bigg( (ab''+a''b)b'+a'(b'^2+2bb'') \bigg)b'c^3\bigg]
\eea
By varying the above full action with respect to each metric component, $a$, $b$ and $c$, and taking the gauge $a=c$, we obtain the following equations:
\bea
&&\phi'^2 + 3 \frac{b''}{b^3} \left( b^2+ 4 G(\phi_i) (1-a^2b'^2) - 8G'(\phi_i)a^2bb' \right) + \frac{12G''(\phi_i)}{b^2}(1-a^2b'^2)= 0 \\ \cr
&&4(-1+a^2b'^2) + (a'^2+aa'')b^2+ ab(7a'b' + 2ab'')  - \frac{12 b^2}{l^2} \cr
&& + \frac{4G(\phi_i)}{b} {\mathcal G}_1 - \frac{4G'(\phi_i)}{b} {\mathcal G}_2 + 4G''(\phi_i)(a^2-a^4b'^2-2a^3a'bb')=0 \\ \cr
&&-1+aa'bb' + a^2b'^2 - \frac{1}{6}a^2b^2 \phi'^2 + \frac{V_{eff}}{3 b^4} - \frac{2 b^2}{l^2} + \frac{4G(\phi_i)}{b} (aa'b' -a^3a'b'^3) \cr
&& - \frac{4G'(\phi_i)}{b} (3a^3a'bb'^2 - aa'b - a^2b' + a^4b'^3) =0
\eea
where
\bea
{\mathcal G}_1 &=& (1-a^2b'^2)(3aa'b'+aa''b) + a'^2b(1-3a^2b'^2)-2a^3a'b b'b'' \\ \cr
{\mathcal G}_2 &=& 6a^2a'^2b^2b' - 3a^2b'(1-a^2b'^2)- 5aa'b(1-3a^2b'^2) + 2a^3b^2(a'b')' + 2a^4bb'b''  \;\;\;\;\;
\eea
Note that differentiating the action with respect to $c(r)$ gave us the Hamiltonian constraint.
\\
The equation of motion for scalar fields is obtained by varying the action with respect to $\phi^i$ and gauging such that $a=c$ as follows:
\bea
\partial_{r}(a^2b^3\partial_{r}\phi_i)
  = &-& \frac{b^3}{2} \partial_i G\; ( R^2  - 4 R_{\mu\nu}R^{\mu\nu} + R_{\mu\nu\alpha\beta} R^{\mu\nu\alpha\beta}) + \frac{1}{b^3} \partial_i(V_{eff})  \nonumber \\
  = &-&12 \partial_i G\;\bigg[ -a'^2b + 3a^2a'^2bb'^2 - a(ab''+2a'b'+a''b) \nonumber \\
    &+& a^3b'\bigg( (ab''+a''b)b'+a'(2b'^2+2bb'') \bigg)\bigg] + \frac{1}{b^3} \partial_i(V_{eff})
\eea

\section{Details for the first order solution}
\label{A2}

From eq.(\ref{phi}), we have
\beq
\partial_r (a^2 b^3 \partial_r \phi_i) = \frac{1}{ b^3} (\partial_i V_{eff}(\phi_i) - \frac{1}{2} b^6 \partial_i V(\phi_i))
\eeq
where $V_{eff}= f^{AB} Q_A Q_B$ and $f^{AB}$ is the inverse of $f_{AB}$.
Then, using eq.(\ref{phi1}), we have
\beq
\phi_i = \phi_{i0} + \delta \phi_i
\eeq
We consider the case where $V$ is constant, and at the zeroth order the above equation is reduced to
\beq
\partial_r (a_0^2 b_0^3 \partial_r \phi_i) = \frac{1}{ b_0^3} \partial_i V_{eff}(\phi_i)
\eeq
Now let us plug $\phi_i = \phi_{i0} + \delta \phi_i$ into the above equation, then we get
\beq
\partial_r (a_0^2 b_0^3 \partial_r \delta \phi_i) = \frac{1}{ b_0^3} \partial_i V_{eff}(\phi_{i0} + \delta \phi_i)
= \frac{1}{ b_0^3} \partial_i^2 V_{eff}(\phi_{i0}) \delta \phi_i
\eeq
Above we have used eq.(\ref{critical}). Now let us define $\partial_i^2 V_{eff}(\phi_{i0})$ to be $\beta_i^2$ and $\delta \phi_i$ to be $\epsilon \phi_{i1}$ and consider the first order in $\epsilon$. We then get eq.(\ref{phi2})
\beq
\label{A5}
\partial_r (a_0^2 b_0^3 \partial_r \phi_{i1}) = \frac{ \beta_i^2}{ b_0^3} \phi_{i1}
\eeq
At the zeroth order, we have
\beq
a_0(r) = (1-\frac{r_H^2}{r^2})\sqrt{1+ \frac{r^2+ 2r_H^2}{l^2}} , \;\;\;\ b_0(r)=r
\eeq
Let us plug these into the above eq.(\ref{A5}), then we get
\beq
\partial_r \left[(1+\frac{r_H}{r})^2 (1-\frac{r_H}{r})^2(1+ \frac{r^2+ 2r_H^2}{l^2}) r^3 \partial_r \phi_{i1}\right] = \frac{ \beta_i^2}{ r^3} \phi_{i1}
\eeq
Now we use eq.(\ref{phi3})
\beq
\phi_{i1} = c_{1i} (1-\frac{r_H}{r})^{\gamma_i}
\eeq
and get
\beq
\partial_r \left[(1+\frac{r_H}{r})^2 (1-\frac{r_H}{r})^2(1+ \frac{r^2+ 2r_H^2}{l^2}) r^3 c_{1i} \gamma_i (1-\frac{r_H}{r})^{\gamma_i - 1}\frac{r_H}{r^2}\right] = \frac{ \beta_i^2}{ r^3} c_{1i} (1-\frac{r_H}{r})^{\gamma_i}
\eeq
For the term of the order $(r-r_H)^{\gamma_i}$, we obtain
\beq
\gamma_i (\gamma_i + 1)(r_H + r_H)^2 (1+ \frac{r_H^2+ 2r_H^2}{l^2}) (1-\frac{r_H}{r})^{\gamma_i} \frac{r_H^2}{r_H^3} = \frac{\beta_i^2}{r_H^3} (1-\frac{r_H}{r})^{\gamma_i}
\eeq
Therefore we obtain eq.(\ref{gamma1})
\beq
\gamma_i (\gamma_i+1) = \frac{\beta_i^2}{4r_H^4} (1+ \frac{3r_H^2}{l^2})^{-1}
\eeq
\\
In a similar way, we can derive the equation (\ref{gamma2}) using (\ref{b}) and (\ref{a}).

\section{Details for the zeroth order Gauss-Bonnet solution}
The Hamiltonian constraint given in  (\ref{eqgb3}) is
\bea
\label{C1}
&&-1+aa'bb' + a^2b'^2 - \frac{1}{6}a^2b^2 \phi'^2 + \frac{V_{eff}}{3 b^4} - \frac{2 b^2}{l^2} + \frac{4G(\phi_i)}{b} (aa'b' -a^3a'b'^3) \cr
&& - \frac{4G'(\phi_i)}{b} (3a^3a'bb'^2 - aa'b - a^2b' + a^4b'^3) = 0
\eea
For the zeroth order solution, we consider $b_0=r$ and constant $\phi_i = \phi_{i0}$. Then the above equation becomes 
\beq
\label{C2}
-1 + ra_0a_0' + a_0^2 + \frac{V_{eff}(\phi_{i0})}{3 r^4} - \frac{2 r^2}{l^2} + \frac{4G(\phi_{i0})}{r} (a_0a_0' -a_0^3a_0') = 0
\eeq
We want the solution for $a_0^2$ to have a double horizon. The double horizon condition determines 
$V_{eff}(\phi_{i0})$ as follows:
\beq
\label{C3}
V_{eff}(\phi_{i0}) = 3 r_H^4 (1 + \frac{2r_H^2}{l^2})
\eeq
Multiplying (\ref{C2}) by $G_0 r$ and plugging $V_{eff}(\phi_{i0})$ into it, we can write (\ref{C2}) as follows:
\beq
\label{C4}
\frac{r^3}{4} + \frac{G_0 r_H^4(1 + \frac{2r_H^2}{l^2})}{r^3} - 2\bigg(G_0 a_0^2 - G_0 - \frac{r^2}{4}\bigg)\bigg(G_0 (a_0^2)' - \frac{r}{2}\bigg) - \frac{2 G_0 r^3}{l^2} = 0
\eeq
which can be easily integrated into the form:
\beq
\label{C5}
\frac{r^4}{16} - \frac{G_0 r_H^4(1 + \frac{2r_H^2}{l^2})}{2 r^2} - \bigg(G_0 a_0^2 - G_0 - \frac{r^2}{4}\bigg)^2 - \frac{G_0 r^4}{2 l^2} + C = 0
\eeq
with integration constant $C$.
Then again requiring the degenerate horizon condition, we set the integration constant C:
\beq
\label{C6}
C = G_0 (G_0 + r_H^2 + \frac{3 r_H^4}{2 l^2})
\eeq
and rearranging (\ref{C5}) we get
\beq
\label{C7}
\bigg(G_0 a_0^2 - G_0 - \frac{r^2}{4}\bigg)^2 = \frac{r^4}{16} \bigg(1 - \frac{8 G_0 r_H^4(1 + \frac{2r_H^2}{l^2})}{r^6} - \frac{8 G_0}{l^2} + \frac{16 G_0 (G_0 + r_H^2 + \frac{3 r_H^4}{2 l^2})}{r^4}\bigg)
\eeq
In obtaining $a_0^2$, we can see that there are two branches $\pm$. But we want to obtain the 
 AdS-Reissner-Nordstrom solution in the limit $G_0 \rightarrow 0$, which makes us choose the $-$ branch. Thus we get 
 (\ref{a}):
\bea
a^2_0 = 1 + \frac{r^2}{4 G_0} -\frac{r^2}{4G_0} \sqrt{1-\frac{8 G_0}{l^2} + \frac{16G_0 (G_0 + r_H^2 + \frac{3r_H^4}{2l^2})}{r^4} -\frac{8G_0r_H^4(1+\frac{2r_H^2}{l^2})}{r^6}}
\eea

\end{document}